\documentclass[12pt]{article}
\usepackage{epsfig}
\usepackage{graphicx}
\usepackage{a4}
\usepackage{latexsym}
\usepackage{cite}

\usepackage{color}
\usepackage{colordvi}

\usepackage{pslatex}
\usepackage[latin1]{inputenc}
\usepackage[T1]{fontenc}

\newcommand{\as}{\alpha_{\rm s}}

\def\MSbar{\overline{\mathrm{MS}}}
\def\ep{\epsilon}
\def\z#1{{\zeta_{#1}}}
\def\ca{{C^{}_A}}
\def\cf{{C^{}_F}}
\def\tf{{T^{}_F}}
\def\nf{{n^{}_{\! f}}}
\def\nl{{n^{}_{\! l}}}
\def\nh{{n^{}_{\! h}}}

\def\lm{\mathrm{L}_m}
\def\ls{\mathrm{L}_s}
\def\lx{\mathrm{L}_x}
\def\ly{\mathrm{L}_y}
\def\s#1#2{\mathrm{S}_{#1,#2}(x)}
\def\li#1{\mathrm{Li}_{#1}(x)}
\newcommand{\ebrk}{\nonumber \\ &&}
\newcommand{\brk}{\right. \nonumber \\ && \left.}
\newcommand{\ibrk}{\right. \right. \nonumber \\ && \left. \left.}
\newcommand{\iibrk}{\right. \right. \right. \nonumber \\ && \left. \left. \left.}

\begin{document}
\setlength{\parskip}{0.2cm} \setlength{\baselineskip}{0.55cm}

\begin{titlepage}
\noindent
DESY 07-064 \hfill {\tt arXiv:0705.1975v1}\\
SFB/CPP-07-19 \\
May 2007 \\
\vspace{1.5cm}
\begin{center}
\LARGE {\bf
Heavy-quark production \\[0.5ex]
in massless quark scattering at two loops in QCD
}\\
\vspace{2.2cm}
\large
M. Czakon$^{a,b}$, A. Mitov$^{c}$ and S. Moch$^{c}$ \\
\vspace{1.4cm}
\normalsize
{\it
$^{a}$Institut f\"ur Theoretische Physik und Astrophysik, Universit\"at W\"urzburg \\[0.5ex]
Am Hubland, D-97074 W\"urzburg, Germany \\[.5cm]
$^{b}$Institute of Nuclear Physics, NCSR ``DEMOKRITOS''\\
15310 Athens, Greece \\[.5cm]
$^{c}$Deutsches Elektronensynchrotron DESY \\[0.5ex]
Platanenallee 6, D--15738 Zeuthen, Germany}
\vfill
\large {\bf Abstract}
\vspace{-0.2cm}
\end{center}
We present the two-loop virtual QCD corrections to the production of
heavy quarks in the quark--anti-quark--annihilation channel in the
limit when all kinematical invariants are large compared to the mass
of the heavy quark. Our result
is exact up to terms suppressed by powers of the heavy-quark mass.
The derivation is based on a simple relation between massless and
massive scattering amplitudes in gauge theories proposed recently by
two of the authors as well as a direct calculation of the massive amplitude
at two loops. The results presented here form an important part of
the next-to-next-to-leading order QCD contributions to 
heavy-quark production in hadron-hadron collisions.
\\
\vspace{2.0cm}
\end{titlepage}

\newpage

%
% -----------------------------------------------------------------------------
%

The hadro-production of heavy quarks, especially of top-quarks, is an important process
at hadron colliders.
Thus far, the Tevatron has provided us with a wealth of information on the top-quark,
most prominently with a precise measurement of its mass.
In the future, the LHC is expected to accumulate very high statistics for the production
of $t\bar{t}$-pairs, approximately $8 \cdot 10^6$ events per year in the initial
low luminosity run~\cite{cms:2006tdr}.
At the LHC the uses of $t\bar{t}$-pairs are e.g. for energy scale calibration,
for background estimates and last but not least, for precision measurements of
Standard Model parameters.

The present and in particular the anticipated future experimental precision
on heavy-quark hadro-production requires theory predictions to include radiative corrections
in Quantum Chromodynamics (QCD) beyond the next-to-leading order (NLO).
For instance, at the LHC the total cross section for $t {\bar t}$-production at NLO in QCD
is accurate to ${\cal O}(15\%)$ only, which has to be contrasted with expected
precision for the top-mass measurement of ${\cal O}(1 {\rm{GeV}})$.
Another important process is the production of $b$-quarks at
moderate to large transverse momentum. NLO QCD corrections for
heavy-quark hadro-production have been known since long, see e.g.
Refs.~\cite{Nason:1988xz,Nason:1989zy,Beenakker:1989bq,Beenakker:1991ma,Mangano:1992jk,Korner:2002hy,Bernreuther:2004jv}.
Knowledge of the radiative QCD corrections at next-to-next-to-leading order (NNLO) will certainly improve the
stability of theory predictions with respect to scale variations and provide a match with precise parton evolution at
NNLO~\cite{Moch:2004pa,Vogt:2004mw}.

In this letter, we present results for the virtual QCD corrections
at two loops for the pair-production of heavy quarks in the $q {\bar
q}$-annihilation channel. To be precise, we calculate the
interference of the two-loop amplitude with the Born one and we work
in the limit of fixed scattering angle and high energy, where all
kinematic invariants are large compared to the heavy-quark mass. Thus,
our result contains all logarithms in the heavy quark mass as well
as all constant contributions (i.e. the mass-independent terms).
Throughout this letter we neglect power corrections in the
heavy-quark mass.

In our calculation we employ two different methods. On the one hand, we
apply a generalization of the infrared factorization formula for
massless QCD amplitudes~\cite{Catani:1998bh,Sterman:2002qn} to the
case of massive partons~\cite{Mitov:2006xs}. In essence, it results
in an extremely simple {\it universal} multiplicative relation
between a massive QCD amplitude in the small-mass limit and its
massless version~\cite{Mitov:2006xs}. In this way, we can largely
use for our derivation the results of Ref.~\cite{Anastasiou:2000kg},
where the NNLO QCD corrections to massless quark-quark scattering
(i.e. $q{\bar q} \to q^\prime {\bar q}^\prime$) have been computed.
On the other hand, we perform a direct calculation of the relevant
Feynman diagrams in the massive case followed by a subsequent
expansion in the small-mass limit. As an added benefit, this
approach provides us with a non-trivial, albeit complicated, check
of the massless results. Moreover, it makes it possible to
systematically calculate power corrections in the mass, which can
improve the convergence of the small-mass expansion. This is
certainly relevant in the case of the top-quark pair-production at
the LHC (less so, perhaps, for $b$-quark production).

We would like to emphasize that the agreement between our prediction
based on the approach of Ref.~\cite{Mitov:2006xs} and our direct
calculation constitutes the first non-trivial check of this
factorization approach at two loops. The formalism of
Ref.~\cite{Mitov:2006xs} has recently been also applied to the
re-derivation of the two-loop QED corrections to Bhabha
scattering~\cite{Becher:2007cu} confirming
earlier results on the two-loop radiative photonic corrections~\cite{%
Bern:2000ie,Glover:2001ev,Penin:2005kf,Penin:2005eh} in that
process.

%
% -----------------------------------------------------------------------------
%
\subsection*{Setting the stage}

The pair-production of heavy quarks in the $q {\bar q}$-annihilation channel
corresponds to the scattering process,
\begin{equation}
\label{eq:qqQQ}
q(p_1) + {\bar q}(p_2) \:\:\rightarrow\:\: Q(p_3,m) + {\bar Q}(p_4,m) \, ,
\end{equation}
where $p_i$ denote the quark momenta and $m$ the mass of the heavy quark.
Energy-momentum conservation implies
\begin{equation}
\label{eq:engmom}
p_1^\mu+p_2^\mu = p_3^\mu+p_4^\mu \, .
\end{equation}
Following the notation of Ref.~\cite{Anastasiou:2000kg} we consider
the scattering amplitude ${\cal M}$ for the process~(\ref{eq:qqQQ})
at fixed values of the external parton momenta $p_i$, thus $p_1^2 =
p_2^2 = 0$ and $p_3^2 = p_4^2 = m^2$. It may be written as a
series expansion in the strong coupling $\as$,
\begin{eqnarray}
  \label{eq:Mexp}
  | {\cal M} \rangle
  & = &
  4 \pi \as \biggl[
  | {\cal M}^{(0)} \rangle
  + \biggl( {\as \over 2 \pi} \biggr) | {\cal M}^{(1)} \rangle
  + \biggl( {\as \over 2 \pi} \biggr)^2 | {\cal M}^{(2)} \rangle
  + {\cal O}(\as^3)
  \biggr]
\, ,
\end{eqnarray}
where we define the expansion coefficients in powers of $\as(\mu^2)
/ (2\pi)$ with $\mu$ being the renormalization scale. We work in
conventional dimensional regularization, $d=4-2 \ep$, in the
$\MSbar$-scheme for the coupling constant renormalization. The heavy
mass $m$ on the other hand is always taken to be the pole mass.

We explicitly relate the bare (unrenormalized) coupling $\as^{\rm{b}}$
to the renormalized coupling $\as$ by
\begin{eqnarray}
\label{eq:alpha-s-renorm}
\as^{\rm{b}} S_\epsilon \: = \: \as
\biggl[
   1
   - {\beta_0 \over \epsilon} \biggl( {\as \over 2 \pi} \biggr)
   + \left(
     {\beta_0^2 \over \epsilon^2}
     - {1 \over 2} {\beta_1 \over \epsilon}
     \right) \biggl( {\as \over 2 \pi} \biggr)^2
  + {\cal O}(\as^3)
  \biggr]
\, ,
\end{eqnarray}
where we put the factor $S_\epsilon=(4 \pi)^\ep \exp(-\ep \gamma_{\rm E}) = 1$
for simplicity and $\beta$ is the QCD $\beta$-function
\cite{vanRitbergen:1997va,Czakon:2004bu}
\begin{eqnarray}
\label{eq:betafct}
\beta_0 = {11 \over 6}\*\ca - {2 \over 3}\*\tf\*\nf \, ,
\qquad
\beta_1 = {17 \over 6}\*\ca^2 - {5 \over 3}\*\ca\*\tf\*\nf - \cf\*\tf\*\nf \, .
\end{eqnarray}
As we work in a general non-Abelian ${\rm{SU}}(N)$-gauge theory we set
$\ca = N$, $\cf = (N^2-1)/2\*N$ and $\tf = 1/2$.
Throughout this letter, $N$ denotes the number of colors and
$\nf$ the total number of flavors, which is the sum of
$\nl$ light and $\nh$ heavy quarks.

The squared amplitude for the process~(\ref{eq:qqQQ})
summed over spins and colors is a function of the
Mandelstam variables $s$, $t$ and $u$ given by
\begin{equation}
\label{eq:Mandelstam}
s = (p_1+p_2)^2\, , \qquad
t  = (p_1-p_3)^2 - m^2\, , \qquad
u  = (p_1-p_4)^2 - m^2\, .
\end{equation}
Then it is convenient to define the function ${\cal A}(\epsilon, m, s, t, \mu)$
for the spin and color averaged amplitudes as
\begin{eqnarray}
\label{eq:Msqrd}
\overline{\sum |{\cal M}({q + \bar{q} \to  Q + \bar{Q}} )|^2}
&=&
{1 \over 4 N^2}\, {\cal A}(\epsilon, m, s, t, \mu)
\, ,
\end{eqnarray}
which has a perturbative expansion similar to Eq.~(\ref{eq:Mexp}),
\begin{equation}
\label{eq:Aexp}
{\cal A}(\epsilon, m, s, t, \mu) = 16 \pi^2 \as^2
\left[
  {\cal A}^4
  + \biggl( {\as \over 2 \pi} \biggr) {\cal A}^6
  + \biggl( {\as \over 2 \pi} \biggr)^2 {\cal A}^8
  + {\cal O}(\as^{3})
\right]
\, .
\end{equation}
In terms of the amplitudes the expansion coefficients in Eq.~(\ref{eq:Aexp})
may be expressed as
\begin{eqnarray}
\label{eq:A4def}
{\cal A}^4 &=&
\langle {\cal M}^{(0)} | {\cal M}^{(0)} \rangle \equiv 2(N^2-1) \left(\frac{t^2+u^2}{s^2} - \epsilon \right)
+ {\cal O}(m)\, , \\
\label{eq:A6def}
{\cal A}^6 &=& \left(
\langle {\cal M}^{(0)} | {\cal M}^{(1)} \rangle + \langle {\cal M}^{(1)} | {\cal M}^{(0)} \rangle
\right)\, , \\
\label{eq:A8def}
{\cal A}^8 &=& \left(
\langle {\cal M}^{(1)} | {\cal M}^{(1)} \rangle
+ \langle {\cal M}^{(0)} | {\cal M}^{(2)} \rangle + \langle {\cal M}^{(2)} | {\cal M}^{(0)} \rangle
\right)\, ,
\end{eqnarray}
where we have neglected powers in the heavy-quark mass $m$ in ${\cal
A}^4$. Expressions for ${\cal A}^6$ with the complete heavy-quark
mass dependence using dimensional regularization can be obtained
e.g. from Ref.~\cite{Korner:2002hy,Bernreuther:2004jv}. The
loop-by-loop contribution $\langle {\cal M}^{(1)} | {\cal M}^{(1)}
\rangle$ in dimensional regularization in ${\cal A}^8$ and also with
the full heavy-quark mass dependence can be computed with the help
of Ref.~\cite{Korner:2005rg}. In this letter, we provide for the
first time the real part of $\langle {\cal M}^{(0)} | {\cal M}^{(2)}
\rangle$ up to powers ${\cal O}(m)$ in the heavy-quark mass $m$.

%
% -----------------------------------------------------------------------------
%
\subsection*{Massive amplitudes from QCD factorization}

Let us briefly recall the key features of Ref.~\cite{Mitov:2006xs}
to calculate loop amplitudes with massive partons from massless
ones. The QCD factorization approach rests on the fact that a
massive amplitude ${\cal M}^{{\rm[p]},(m)}$ for any given physical
process shares essential properties in the small-mass limit with
the corresponding massless amplitude ${\cal M}^{{\rm[p]},(m=0)}$. The
latter one, ${\cal M}^{{\rm[p]},(m=0)}$, generally displays two
types of singularities, soft and collinear, related to the emission
of gluons with vanishing energy and to collinear parton radiation
off massless hard partons, respectively. These appear explicitly as
factorizing poles in $\ep$ in dimensional regularization after the
usual ultraviolet renormalization is performed. In the former case,
the soft singularities remain in ${\cal M}^{{\rm[p]},(m)}$ as single
poles in $\ep$ while some of the collinear singularities are now
screened by the mass $m$ of the heavy fields, which gives rise to
a logarithmic dependence on $m$, see e.g. Ref.~\cite{Catani:2000ef}.

Thus, in the small-mass limit the differences between a massless and
a massive amplitude can be thought of as a mere change in the
regularization scheme. As an upshot, QCD factorization provides a
direct relation between ${\cal M}^{{\rm[p]},(m)}$ and ${\cal
M}^{{\rm[p]},(m=0)}$ which can be cast in the remarkably simple and
suggestive relation
\begin{eqnarray}
\label{eq:Mm-M0}
{\cal M}^{{\rm[p]},(m)} &=&
\prod_{i\in\ \{{\rm all}\ {\rm legs}\}}\,
  \left(
    Z^{(m\vert0)}_{[i]}
  \right)^{1 \over 2}\,
  \times\
{\cal M}^{{\rm[p]},(m=0)}\, .
\end{eqnarray}

The function $Z^{(m\vert 0)}$ is process independent and depends
only on the external parton, i.e. quarks in the case at hand. For
external massive quarks $Q$ it is defined as the ratio of the
on-shell heavy-quark form factor and the massless on-shell one, both
being known~\cite{Bernreuther:2004ih,Moch:2005id,Gehrmann:2005pd} to
sufficient orders in $\as$ and powers of $\ep$. An explicit
expression for
\begin{equation}
Z^{(m\vert0)}_{[Q]} \, = \, 1 + \sum\limits_{j=1}^{\infty} \left(
{\as \over 2 \pi} \right)^j\, Z^{(j)} \, ,
\end{equation}
up to two loops is given
\footnote{Note the different normalization ${\as /  (4 \pi)}$ of the
coupling used in that reference.}
in Ref.~\cite{Mitov:2006xs}. Exploiting the full predictive power of
the relation Eq.~(\ref{eq:Mm-M0}) and applying it to the process
Eq.~(\ref{eq:qqQQ}) we get
\begin{equation}
\label{eq:A8mtoA80} 2 {\rm Re}\, \langle {\cal M}^{(0)} | {\cal
M}^{(2)} \rangle^{(m)} = 2 {\rm Re}\, \langle {\cal M}^{(0)} |
{\cal M}^{(2)} \rangle^{(m=0)}
+ Z^{(1)} {\cal A}^{6,(m=0)}\ +\ 2 Z^{(2)} {\cal
A}^{4,(m=0)} \, ,
\end{equation}
which assumes the hierarchy of scales $m^2 \ll s,t,u$ , i.e. we
neglect terms ${\cal O}(m)$. Eq.~(\ref{eq:A8mtoA80}) predicts the
complete real part of the squared amplitude $\langle {\cal M}^{(0)}
| {\cal M}^{(2)} \rangle$ except for those terms, which are linear
in $\nh$, i.e. the number of heavy quarks. These two-loop
contributions have been excluded explicitly from the definition
\cite{Mitov:2006xs} of $Z^{(m\vert 0)}$, as one needs additional
process dependent terms for their description. Their incorporation
for the case of Bhabha scattering was presented in
Ref.~\cite{Becher:2007cu} in agreement with the direct 
calculation~\cite{Actis:2007gi}.

%
% -----------------------------------------------------------------------------
%
\subsection*{Direct calculation of the massive amplitude}

An alternative method is the direct calculation of all necessary
massive Feynman diagrams together with an expansion in the small
mass. The advantage of this approach is an independent check of
Eq.~(\ref{eq:A8mtoA80}) as well as of the corresponding massless
results. Moreover, it also allows for a relatively easy access to
all heavy-quark loop corrections. Presently, these have not been
obtained from the QCD factorization method, since they are related
to the process dependent contributions.

We performed our computation using the DiaGen/IdSolver system of one
of the authors (M.C.). After the diagram generation phase, all the
integrals have been reduced to a set of 145 masters with the help of
the Laporta algorithm \cite{Laporta:2001dd} extended by topological
symmetry properties, where, however, integrals which are related by
a $t \leftrightarrow u$ exchange of the Mandelstam variables are
considered as being independent. The evaluation of the masters
proceeded similarly to the methodology developed in
Ref.~\cite{Czakon:2004wm,Czakon:2006pa}. The main idea was to construct
Mellin-Barnes \cite{Smirnov:1999gc,Tausk:1999vh} representations for all the
integrals, followed by a subsequent analytic continuation with the MB package
\cite{Czakon:2005rk} and an expansion in the mass by closing
contours. The resulting integrals have then been transformed into
series representations, some two-fold, and resummed with the help of
XSummer~\cite{Moch:2005uc} in the cases of non-trivial dependence on
the kinematic variables. Some constants, though, were not given by
harmonic series and have been computed with the help of the PSLQ
algorithm~\cite{pslq:1992}. Needless to say that all of the above
steps were performed fully automatically.

\begin{figure}[ht]
  \parbox[t]{9cm}{\center \parbox[c][4cm][c]{8cm}{
    \epsfig{file=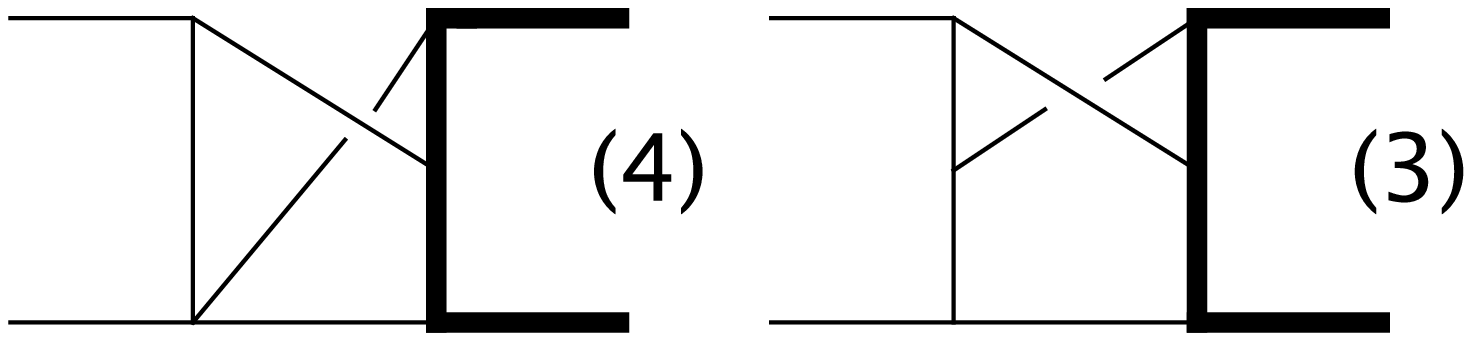,width=8cm}}
    \caption{\label{fig:np}
      The most complicated non-planar topologies, together with the multiplicities of
      the respective master integrals for $q{\bar q} \to Q {\bar Q}$
      scattering. The thick lines are massive.
    }}
  \hspace{1cm}
  \parbox[t]{6cm}{\center \parbox[c][4cm][c]{5cm}{
    \epsfig{file=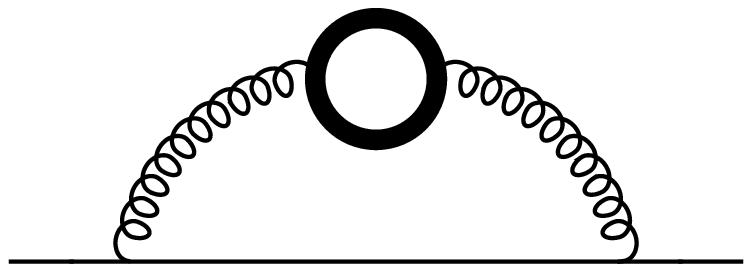,width=5cm}}
    \caption{\label{fig:z2}
     A heavy-quark loop contribution to the wave-function renormalization of
     the light quarks.}}
\end{figure}
In our calculation, we have only been able to obtain the leading
color term and the full dependence on the number of light and heavy
fermion species. The reason is that the terms subleading in color
contain contributions from non-planar graphs. Fig.~\ref{fig:np}
shows the bottleneck cases with 6- and 8-fold Mellin-Barnes
representations, for the six and seven liners respectively. The
representations have been obtained similarly as in
Ref.~\cite{Tausk:1999vh}, i.e. from the Feynman-parametric
representation of the two-loop graphs and not by integrating
loop-by-loop, the strategy adopted for planars. In this way, the
mass expansion generates integrals that are at worst as complex as
the massless ones, plus some new terms that behave as $1/\sqrt{m^2}$
and must cancel in the complete result. The sheer amount of
remaining difficult 4-fold integrals turned out to be presently
intractable with the developed software.

Finally, as mentioned above, we renormalized the mass of the heavy quark and the
external wave functions in the on-shell scheme
(besides the $\MSbar$ renormalization of the coupling constant).
The former results are now known up to three-loop level from Ref.~\cite{Melnikov:2000qh,Melnikov:2000zc}.
A peculiarity of the light quark states is that they obtain contributions to
the wave-function renormalization constants from heavy-quark loops, see
Fig.~\ref{fig:z2}.
In fact, we have
\begin{equation}
\label{eq:Z2wavefct}
Z_2 = 1+ \left(\frac{\alpha_s}{2\pi}\right)^2 \cf \tf \nh
\left[
  \frac{1}{4\epsilon}
  - \frac{1}{2} \log \left( \frac{m^2}{\mu^2} \right)
  - \frac{5}{24}
\right]\, .
\end{equation}
%

%
% -----------------------------------------------------------------------------
%
\subsection*{Results}

We are now in a position to present our result
for $q{\bar q} \to Q {\bar Q}$ scattering for the
interference of the two-loop and Born amplitude,
\begin{eqnarray}
  \label{eq:ReM0x2}
{\lefteqn{
  2 {\rm Re}\, \langle {\cal M}^{(0)} | {\cal M}^{(2)} \rangle = }}
\\
& &
2 (N^2-1) \biggl(
N^2 A + B  + {1 \over N^2} C
+ N \nl D_l + N \nh D_h
+ {\nl \over N} E_l + {\nh \over N} E_h + (\nl+ \nh)^2 F
\biggr)
\, ,
\nonumber
\end{eqnarray}
which we choose to express by grouping terms according to the power
of the number of colors $N$ and the numbers of $\nl$ light and
$\nh=1$ heavy quarks with $\nf=\nl+\nh$ total flavors. As detailed
above, the coefficients $A$, $D_l$, $E_l$ and $F$ have been computed
with both methods, i.e. by employing our universal multiplicative
relation~(\ref{eq:Mm-M0}) between the massive and massless
amplitudes as well as a direct evaluation of the loop integrals in
the small-mass expansion. The terms linear in $\nh$, $D_h$ and $E_h$
could also be easily obtained from a direct Feynman diagram
calculation, while for $B$ and $C$ the approach based on
factorization proved more powerful.

As the only dimensionless kinematic variable in
our problem, we choose $x=-t/s$ and keep the dependence on the renormalization
scale, $\mu$, explicit. We also introduce the following compact notation
\begin{equation}
\lm = \log\left( \frac{m^2}{s} \right)\, , \;\;\;\;
\ls = \log\left( \frac{s}{\mu^2} \right)\, , \;\;\;\;
\lx = \log\left( x \right)\, , \;\;\;\;
\ly = \log\left( 1-x \right)\, .
\end{equation}

The different components now read
{\small{
%
%%START
\begin{eqnarray}
A &=&
\frac{1}{\epsilon^4}\left\{
\frac{x^2}{2}-\frac{x}{2}+\frac{1}{4}
\right\}
+
\frac{1}{\epsilon^3}\left\{
\lm\left[
x^2-x+\frac{1}{2}
\right]
+
\ls\left[
-x^2+x-\frac{1}{2}
\right]
+
\frac{21 x^2}{4}-\frac{21 x}{4}
+\lx \left(-2 x^2
\ibrk
+2
   x-1\right)+\frac{19}{8}
\right\}
+
\frac{1}{\epsilon^2}\left\{
\lm \ls\left[
-2 x^2+2 x-1
\right]
+
\ls^2\left[
x^2-x+\frac{1}{2}
\right]
+
\lm\left[
\frac{29 x^2}{6}-\frac{29 x}{6}
\ibrk
+\lx \left(-2 x^2+2
   x-1\right)+\frac{23}{12}
\right]
+
\ls\left[
-\frac{19 x^2}{6}+\frac{19 x}{6}+\lx \left(4 x^2-4
   x+2\right)
-\frac{13}{12}
\right]
+
\left(2 x^2
\ibrk
-\frac{5 x}{2}+\frac{5}{4}\right) \lx^2+\left(-\frac{26
   x^2}{3}+\frac{55 x}{6}-\frac{23}{6}\right) \lx+\frac{173
   x^2}{72}-\frac{173 x}{72}+\pi ^2
   \left(-\frac{x^2}{6}+\frac{x}{6}-\frac{1}{12}\right)
\brk
-\frac{205}{144}
\right\}
+
\frac{1}{\epsilon}\left\{
\lm^3\left[
-\frac{x^2}{3}+\frac{x}{3}-\frac{1}{6}
\right]
+
\lm \ls^2\left[
2 x^2-2 x+1
\right]
+
\ls^3\left[
-\frac{2 x^2}{3}+\frac{2 x}{3}-\frac{1}{3}
\right]
\brk
+
\lm^2\left[
-x^2+x+\lx \left(x^2-x+\frac{1}{2}\right)-\frac{1}{2}
\right]
+
\lm \ls\left[
-\frac{7 x^2}{3}+\frac{7 x}{3}+\lx \left(4 x^2-4 x+2\right)-\frac{1}{6}
\right]
\brk
+
\ls^2\left[
-\frac{x^2}{2}+\frac{x}{2}+\lx \left(-4 x^2+4 x-2\right)-\frac{3}{4}
\right]
+
\lm\left[
\left(\frac{1}{4}-\frac{x}{2}\right) \lx^2+\left(\frac{3
   x}{2}-x^2\right) \lx-\frac{47 x^2}{12}
\ibrk
+\frac{47 x}{12}-\frac{35}{8}
\right]
+
\ls\left[
\left(-4 x^2+5 x-\frac{5}{2}\right) \lx^2+\left(\frac{8 x^2}{3}-\frac{11
   x}{3}+\frac{1}{3}\right) \lx+\frac{487 x^2}{36}-\frac{487 x}{36}
\ibrk
+\pi
   ^2 \left(\frac{x^2}{3}-\frac{x}{3}+\frac{1}{6}\right)+\frac{601}{72}
\right]
+
\left(\frac{x^2}{3}+x-\frac{1}{2}\right) \lx^3+\left(-\frac{5
   x}{2}+\ly
   \left(-x^2+\frac{x}{2}-\frac{1}{4}\right)+\frac{3}{4}\right)
   \lx^2
\brk
+\li2 \left(-2 x^2+x-\frac{1}{2}\right)
   \lx+\left(\frac{43 x^2}{3}-\frac{151 x}{12}+\pi ^2 \left(\frac{4
   x^2}{3}-\frac{5 x}{6}+\frac{5}{12}\right)+10\right) \lx
\brk
-\frac{9907
   x^2}{432}+\frac{9907 x}{432}+\pi ^2 \left(-\frac{23 x^2}{72}+\frac{5
   x}{72}+\frac{25}{144}\right)+\li3 \left(2
   x^2-x+\frac{1}{2}\right)+\left(-\frac{23 x^2}{6}
\ibrk
+\frac{17
   x}{6}-\frac{17}{12}\right) \z3-\frac{10945}{864}
\right\}
+
\lm^4\left[
\frac{x^2}{4}-\frac{x}{4}+\frac{1}{8}
\right]
+
\lm^3 \ls\left[
\frac{2 x^2}{3}-\frac{2 x}{3}+\frac{1}{3}
\right]
+
\lm \ls^3\left[
-\frac{4 x^2}{3}
\brk
+\frac{4 x}{3}-\frac{2}{3}
\right]
+
\ls^4\left[
\frac{x^2}{3}-\frac{x}{3}+\frac{1}{6}
\right]
+
\lm^3\left[
-\frac{11 x^2}{18}+\frac{11 x}{18}+\lx
   \left(-\frac{x^2}{3}+\frac{x}{3}-\frac{1}{6}\right)-\frac{5}{36}
\right]
\ebrk
+
\lm^2 \ls\left[
-\frac{5 x^2}{3}+\frac{5 x}{3}+\lx \left(-2 x^2+2 x-1\right)-\frac{5}{6}
\right]
+
\lm \ls^2\left[
-\frac{4 x^2}{3}+\frac{4 x}{3}+\lx \left(-4 x^2+4 x-2\right)
\brk
-\frac{5}{3}
\right]
+
\ls^3\left[
\frac{14 x^2}{9}-\frac{14 x}{9}+\lx \left(\frac{8 x^2}{3}-\frac{8
   x}{3}+\frac{4}{3}\right)+\frac{10}{9}
\right]
+
\lm^2\left[
\left(\frac{x}{4}-\frac{1}{8}\right) \lx^2+\left(\frac{x^2}{2}-\frac{3
   x}{4}\right) \lx
\brk
+\frac{247 x^2}{36}-\frac{247 x}{36}+\frac{283}{72}
\right]
+
\lm \ls\left[
\left(x-\frac{1}{2}\right) \lx^2+\left(2 x^2-3 x\right)
   \lx+\frac{23 x^2}{2}-\frac{23 x}{2}+\frac{83}{12}
\right]
\ebrk
+
\ls^2\left[
\left(4 x^2-5 x+\frac{5}{2}\right) \lx^2+\left(\frac{14 x^2}{3}-\frac{11
   x}{3}+\frac{10}{3}\right) \lx-\frac{37 x^2}{4}+\frac{37 x}{4}+\pi ^2
   \left(-\frac{x^2}{3}+\frac{x}{3}-\frac{1}{6}\right)
\brk
-\frac{35}{8}
\right]
+
\lm\left[
\frac{x^2 \lx^3}{3}+\left(-\frac{x}{2}+\ly
   \left(-x^2+\frac{x}{2}-\frac{1}{4}\right)+\frac{1}{4}\right)
   \lx^2+\li2 \left(-2 x^2+x-\frac{1}{2}\right) \lx
\brk
+\left(-4
   x^2+\frac{19 x}{4}+\pi ^2 \left(x^2-\frac{x}{2}+\frac{1}{4}\right)-2\right)
   \lx-\frac{781 x^2}{72}+\frac{781 x}{72}+\pi ^2 \left(-\frac{7
   x^2}{12}+\frac{x}{3}-\frac{1}{24}\right)
\brk
+\li3 \left(2
   x^2-x+\frac{1}{2}\right)+\left(\frac{7 x^2}{3}-\frac{10
   x}{3}+\frac{5}{3}\right) \z3-\frac{499}{144}
\right]
+
\ls\left[
\left(-\frac{2 x^2}{3}-2 x+1\right) \lx^3
\brk
+\left(\frac{4 x}{3}+\ly
   \left(2 x^2-x+\frac{1}{2}\right)+\frac{1}{3}\right) \lx^2+\li2
   \left(4 x^2-2 x+1\right) \lx+\left(-\frac{86 x^2}{3}+\frac{173
   x}{6}
\ibrk
+\pi ^2 \left(-\frac{8 x^2}{3}+\frac{5
   x}{3}-\frac{5}{6}\right)-\frac{49}{3}\right) \lx+\frac{2003
   x^2}{216}-\frac{2003 x}{216}+\li3 \left(-4 x^2+2 x-1\right)
\brk
+\pi ^2
   \left(-\frac{43 x^2}{36}+\frac{61
   x}{36}-\frac{91}{72}\right)+\left(\frac{23 x^2}{3}-\frac{17
   x}{3}+\frac{17}{6}\right) \z3-\frac{919}{432}
\right]
+
\left(-x^2-\frac{x}{24}+\frac{1}{48}\right) \lx^4
\ebrk
+\left(-\frac{7
   x^2}{18}+\frac{13 x}{12}+\ly \left(\frac{10 x^2}{3}-2
   x+1\right)-\frac{7}{12}\right)
   \lx^3+\left(\left(-\frac{x^2}{2}-\frac{5 x}{4}+\frac{11}{8}\right)
   \ly^2+\left(\frac{7 x^2}{6}
\ibrk
+\frac{13 x}{12}+\frac{13}{12}\right)
   \ly+\frac{101 x}{72}+\pi ^2 \left(-\frac{25 x^2}{6}+\frac{25
   x}{12}-\frac{25}{24}\right)+\frac{3}{1-x}-\frac{617}{144}\right)
   \lx^2+\s12 \left(-2 x^2
\brk
-5 x+\frac{11}{2}\right)
   \lx+\left(\frac{260 x^2}{9}-\frac{139 x}{4}+\pi ^2 \left(-\frac{25
   x^2}{9}+\frac{53 x}{36}-\frac{47}{18}\right)+\ly \pi ^2
   \left(-\frac{x^2}{3}+\frac{11 x}{6}-\frac{17}{12}\right)
\brk
+\left(\frac{52
   x^2}{3}-\frac{31 x}{3}+\frac{31}{6}\right) \z3+\frac{1027}{72}\right)
   \lx+\frac{17845 x^2}{2592}-\frac{17845 x}{2592}+\pi ^4
   \left(-\frac{11 x^2}{80}-\frac{x}{240}+\frac{1}{480}\right)
\ebrk
+\li4
   \left(2 x^2+3 x-\frac{3}{2}\right)+\s22 \left(2 x^2+5
   x-\frac{11}{2}\right)+\pi ^2 \left(\frac{1009 x^2}{432}-\frac{259
   x}{432}+\frac{403}{864}\right)
\ebrk
+\li3 \left(-\frac{7 x^2}{3}-\frac{13
   x}{6}+\lx \left(-8 x^2+2 x-1\right)+\ly \left(2 x^2+5
   x-\frac{11}{2}\right)-\frac{13}{6}\right)+\li2 \left(\left(7
   x^2
\ibrk
-\frac{7 x}{2}+\frac{7}{4}\right) \lx^2+\left(\frac{7
   x^2}{3}+\frac{13 x}{6}+\ly \left(-2 x^2-5
   x+\frac{11}{2}\right)+\frac{13}{6}\right) \lx+\pi ^2
   \left(-\frac{x^2}{3}+\frac{11 x}{6}-\frac{17}{12}\right)\right)
\ebrk
+\ly
   \left(-2 x^2-5 x+\frac{11}{2}\right) \z3+\left(-\frac{14
   x^2}{9}+\frac{55 x}{18}+\frac{149}{36}\right) \z3+\frac{77287}{5184},
\end{eqnarray}

\begin{eqnarray}
B &=&
\frac{1}{\epsilon^4}\left\{
-x^2+x-\frac{1}{2}
\right\}
+
\frac{1}{\epsilon^3}\left\{
\lm\left[
-2 x^2+2 x-1
\right]
+
\ls\left[
2 x^2-2 x+1
\right]
-\frac{31 x^2}{4}+\frac{31 x}{4}+\ly \left(-4 x^2
\ibrk
+4 x-2\right)+\lx
   \left(6 x^2-6 x+3\right)-\frac{27}{8}
\right\}
+
\frac{1}{\epsilon^2}\left\{
\lm \ls\left[
4 x^2-4 x+2
\right]
+
\ls^2\left[
-2 x^2+2 x-1
\right]
\brk
+
\lm\left[
-\frac{47 x^2}{6}+\frac{47 x}{6}+\ly \left(-4 x^2+4 x-2\right)+\lx
   \left(6 x^2-6 x+3\right)-\frac{35}{12}
\right]
+
\ls\left[
\frac{49 x^2}{6}-\frac{49 x}{6}
\ibrk
+\lx \left(-12 x^2+12
   x-6\right)+\ly \left(8 x^2-8 x+4\right)+\frac{37}{12}
\right]
+
\left(-6 x^2+\frac{15 x}{2}-\frac{15}{4}\right) \lx^2+\left(\frac{67
   x^2}{3}
\ibrk
-\frac{143 x}{6}+\ly \left(4 x^2-4
   x+2\right)+\frac{29}{3}\right) \lx-\frac{415 x^2}{36}+\frac{415
   x}{36}+\ly \left(-\frac{52 x^2}{3}+\frac{49
   x}{3}-\frac{20}{3}\right)
\brk
+\pi ^2 \left(\frac{17 x^2}{12}-\frac{17
   x}{12}+\frac{17}{24}\right)+\ly^2 \left(2
   x^2-x+\frac{1}{2}\right)-\frac{17}{9}
\right\}
+
\frac{1}{\epsilon}\left\{
\lm^3\left[
\frac{2 x^2}{3}-\frac{2 x}{3}+\frac{1}{3}
\right]
\brk
+
\lm \ls^2\left[
-4 x^2+4 x-2
\right]
+
\ls^3\left[
\frac{4 x^2}{3}-\frac{4 x}{3}+\frac{2}{3}
\right]
+
\lm^2\left[
2 x^2-2 x+\lx \left(-3 x^2+3 x-\frac{3}{2}\right)
\ibrk
+\ly \left(2
   x^2-2 x+1\right)+1
\right]
+
\lm \ls\left[
\frac{25 x^2}{3}-\frac{25 x}{3}+\lx \left(-12 x^2+12
   x-6\right)+\ly \left(8 x^2-8 x+4\right)
\ibrk
+\frac{13}{6}
\right]
+
\ls^2\left[
-\frac{9 x^2}{2}+\frac{9 x}{2}+\ly \left(-8 x^2+8 x-4\right)+\lx
   \left(12 x^2-12 x+6\right)-\frac{5}{4}
\right]
+
\lm\left[
\left(\frac{3 x}{2}
\iibrk
-\frac{3}{4}\right) \lx^2+\left(3 x^2-\frac{9
   x}{2}\right) \lx-\frac{16 x^2}{3}+\ly^2
   \left(x-\frac{1}{2}\right)+\frac{16 x}{3}+\ly \left(-2
   x^2+x+1\right)+\pi ^2 \left(\frac{7 x^2}{6}
\iibrk
-\frac{7
   x}{6}+\frac{7}{12}\right)+\frac{5}{4}
\right]
+
\ls\left[
\left(12 x^2-15 x+\frac{15}{2}\right) \lx^2+\left(-\frac{46
   x^2}{3}+\frac{55 x}{3}+\ly \left(-8 x^2+8
   x-4\right)
\iibrk
-\frac{14}{3}\right) \lx+\frac{85 x^2}{18}-\frac{85
   x}{18}+\ly^2 \left(-4 x^2+2 x-1\right)+\pi ^2 \left(-\frac{17
   x^2}{6}+\frac{17 x}{6}-\frac{17}{12}\right)+\ly \left(\frac{16
   x^2}{3}
\iibrk
-\frac{10 x}{3}-\frac{4}{3}\right)-\frac{31}{18}
\right]
+
\left(-x^2-3 x+\frac{3}{2}\right) \lx^3+\left(\frac{15 x}{2}+\ly
   \left(3 x^2-\frac{x}{2}+\frac{1}{4}\right)-\frac{9}{4}\right)
   \lx^2
\brk
+\li2 \left(6 x^2-3 x+\frac{3}{2}\right)
   \lx+\left(\left(\frac{1}{2}-x\right) \ly^2-\ly-\frac{50
   x^2}{3}+\frac{137 x}{12}+\pi ^2 \left(-\frac{31 x^2}{3}+\frac{41
   x}{6}-\frac{41}{12}\right)
\ibrk
-15\right) \lx-\frac{419
   x^2}{108}+\frac{419 x}{108}+\ly^2 \left(4
   x-\frac{5}{2}\right)+\li3 \left(-6 x^2+3
   x-\frac{3}{2}\right)+\s12 \left(-4 x^2+6 x-3\right)
\brk
+\ly^3
   \left(\frac{2 x^2}{3}-\frac{7 x}{3}+\frac{7}{6}\right)+\pi ^2
   \left(\frac{263 x^2}{72}-\frac{101 x}{72}-\frac{91}{144}\right)+\ly
   \left(\frac{86 x^2}{3}-\frac{193 x}{6}+\pi ^2 \left(\frac{8
   x^2}{3}-\frac{11
   x}{3}
\iibrk
+\frac{11}{6}\right)+\frac{47}{2}\right)+\left(\frac{91
   x^2}{6}-\frac{73 x}{6}+\frac{73}{12}\right) \z3+\frac{413}{108}
\right\}
+
\lm^4\left[
-\frac{x^2}{2}+\frac{x}{2}-\frac{1}{4}
\right]
+
\lm^3 \ls\left[
-\frac{4 x^2}{3}+\frac{4 x}{3}
\brk
-\frac{2}{3}
\right]
+
\lm \ls^3\left[
\frac{8 x^2}{3}-\frac{8 x}{3}+\frac{4}{3}
\right]
+
\ls^4\left[
-\frac{2 x^2}{3}+\frac{2 x}{3}-\frac{1}{3}
\right]
+
\lm^3\left[
\frac{11 x^2}{18}-\frac{11 x}{18}+\ly \left(-\frac{2 x^2}{3}+\frac{2
   x}{3}
\ibrk
-\frac{1}{3}\right)+\lx
   \left(x^2-x+\frac{1}{2}\right)-\frac{1}{36}
\right]
+
\lm^2 \ls\left[
-\frac{x^2}{3}+\frac{x}{3}+\ly \left(-4 x^2+4 x-2\right)+\lx
   \left(6 x^2-6 x+3\right)
\brk
-\frac{1}{6}
\right]
+
\lm \ls^2\left[
-\frac{14 x^2}{3}+\frac{14 x}{3}+\ly \left(-8 x^2+8 x-4\right)+\lx
   \left(12 x^2-12 x+6\right)-\frac{1}{3}
\right]
+
\ls^3\left[
\frac{16 x^2}{9}
\brk
-\frac{16 x}{9}+\lx \left(-8 x^2+8 x-4\right)+\ly
   \left(\frac{16 x^2}{3}-\frac{16 x}{3}+\frac{8}{3}\right)+\frac{2}{9}
\right]
+
\lm^2\left[
\left(\frac{3}{8}-\frac{3 x}{4}\right) \lx^2+\left(\frac{9 x}{4}
\ibrk
-\frac{3
   x^2}{2}\right) \lx-\frac{157 x^2}{36}+\ly^2
   \left(\frac{1}{4}-\frac{x}{2}\right)+\frac{157 x}{36}+\pi ^2 \left(-\frac{2
   x^2}{3}+\frac{2 x}{3}-\frac{1}{3}\right)+\ly
   \left(x^2-\frac{x}{2}-\frac{1}{2}\right)
\brk
-\frac{229}{72}
\right]
+
\lm \ls\left[
\left(\frac{3}{2}-3 x\right) \lx^2+\left(9 x-6 x^2\right) \lx+7
   x^2+\ly^2 (1-2 x)-7 x+\pi ^2 \left(-\frac{7 x^2}{3}+\frac{7
   x}{3}
\ibrk
-\frac{7}{6}\right)+\ly \left(4 x^2-2 x-2\right)-\frac{2}{3}
\right]
+
\ls^2\left[
\left(-12 x^2+15 x-\frac{15}{2}\right) \lx^2+\left(\frac{2
   x^2}{3}-\frac{11 x}{3}+\ly \left(8 x^2-8
   x
\iibrk
+4\right)-\frac{8}{3}\right) \lx+\frac{40 x^2}{9}-\frac{40 x}{9}+\pi
   ^2 \left(\frac{17 x^2}{6}-\frac{17 x}{6}+\frac{17}{12}\right)+\ly^2
   \left(4 x^2-2 x+1\right)+\ly \left(\frac{28 x^2}{3}-\frac{34
   x}{3}
\ibrk
+\frac{26}{3}\right)+\frac{161}{36}
\right]
+
\lm\left[
-x^2 \lx^3+\left(\frac{3 x}{2}+\ly \left(3 x^2-\frac{3
   x}{2}+\frac{3}{4}\right)-\frac{3}{4}\right) \lx^2+\li2 \left(6
   x^2-3 x
\ibrk
+\frac{3}{2}\right) \lx+\left(12 x^2-\frac{57 x}{4}+\pi ^2
   \left(-3 x^2+\frac{3 x}{2}-\frac{3}{4}\right)+6\right) \lx-\frac{115
   x^2}{9}+\ly^2 \left(x-\frac{1}{2}\right)+\frac{115 x}{9}
\brk
+\li3
   \left(-6 x^2+3 x-\frac{3}{2}\right)+\s12 \left(-4 x^2+6
   x-3\right)+\ly^3 \left(\frac{2 x^2}{3}-\frac{4
   x}{3}+\frac{2}{3}\right)+\pi ^2 \left(\frac{9
   x^2}{4}-x
\ibrk
-\frac{5}{24}\right)+\ly \left(-8 x^2+\frac{13 x}{2}+\pi ^2
   \left(2 x^2-3 x+\frac{3}{2}\right)-\frac{5}{2}\right)+\left(\frac{37
   x^2}{3}-\frac{28 x}{3}+\frac{14}{3}\right) \z3-\frac{67}{18}
\right]
\ebrk
+
\ls\left[
\left(2 x^2+6 x-3\right) \lx^3+\left(-\frac{23 x}{3}+\ly \left(-6
   x^2+x-\frac{1}{2}\right)+\frac{5}{6}\right) \lx^2+\li2
   \left(-12 x^2+6 x-3\right) \lx
\brk
+\left((2 x-1) \ly^2+2
   \ly+\frac{100 x^2}{3}-\frac{181 x}{6}+\pi ^2 \left(\frac{62
   x^2}{3}-\frac{41 x}{3}+\frac{41}{6}\right)+\frac{68}{3}\right)
   \lx-\frac{1957 x^2}{54}+\ly^2 \left(\frac{4}{3}
\ibrk
-\frac{2
   x}{3}\right)+\frac{1957 x}{54}+\ly^3 \left(-\frac{4 x^2}{3}+\frac{14
   x}{3}-\frac{7}{3}\right)+\pi ^2 \left(\frac{67 x^2}{36}-\frac{229
   x}{36}+\frac{421}{72}\right)+\s12 \left(8 x^2-12
   x
\ibrk
+6\right)+\li3 \left(12 x^2-6 x+3\right)+\ly \left(-\frac{172
   x^2}{3}+57 x+\pi ^2 \left(-\frac{16 x^2}{3}+\frac{22
   x}{3}-\frac{11}{3}\right)-\frac{97}{3}\right)
\brk
+\left(-\frac{91
   x^2}{3}+\frac{73 x}{3}-\frac{73}{6}\right) \z3-\frac{553}{27}
\right]
+
\left(3 x^2+\frac{x}{8}-\frac{1}{16}\right)
   \lx^4+\left(-\frac{x^2}{18}-\frac{35 x}{12}+\ly \left(-\frac{32
   x^2}{3}
\ibrk
+\frac{16 x}{3}-\frac{8}{3}\right)+\frac{7}{4}\right)
   \lx^3+\left(\left(\frac{7 x^2}{2}+\frac{5 x}{4}-\frac{17}{8}\right)
   \ly^2+\left(\frac{x^2}{6}-\frac{29 x}{12}-\frac{13}{6}\right)
   \ly+\frac{365 x}{72}+\pi ^2 \left(\frac{95 x^2}{6}
\ibrk
-\frac{103
   x}{12}+\frac{103}{24}\right)-\frac{7}{1-x}+\frac{1027}{144}\right)
   \lx^2+\left(\left(-\frac{2 x^2}{3}+\frac{4 x}{3}-\frac{2}{3}\right)
   \ly^3+\left(\frac{x}{2}-\frac{1}{4}\right) \ly^2+\left(\pi ^2
   \left(-\frac{43 x^2}{3}
\iibrk
+\frac{15 x}{2}-\frac{11}{4}\right)+2\right)
   \ly-\frac{304 x^2}{9}+\frac{153 x}{4}+\pi ^2 \left(-\frac{125
   x^2}{18}+\frac{149 x}{36}+\frac{155}{36}\right)+\left(-54 x^2+45
   x
\ibrk
-\frac{45}{2}\right) \z3-\frac{1181}{72}\right)
   \lx+\frac{38959 x^2}{648}-\frac{38959 x}{648}+\s22 \left(-10
   x^2+25 x-\frac{1}{2}\right)+\li4 \left(-6
   x^2
\brk
-x+\frac{1}{2}\right)+\ly^4 \left(-x^2+\frac{25
   x}{12}-\frac{25}{24}\right)+\ly^3 \left(-\frac{7 x^2}{9}+\frac{2
   x}{9}-\frac{13}{36}\right)+\pi ^4 \left(\frac{77 x^2}{45}-\frac{29
   x}{24}+\frac{101}{240}\right)
\ebrk
+\s13 \left(4 x^2+2
   x-1\right)+\ly^2 \left(-\frac{281 x}{36}+\pi ^2 \left(-4 x^2+\frac{19
   x}{3}-\frac{19}{6}\right)-\frac{163}{72}+\frac{5}{x}\right)
\ebrk
+\s12
   \left(\frac{14 x^2}{3}-\frac{47 x}{3}+\ly \left(8 x^2-8
   x+4\right)+\lx \left(10 x^2-9
   x-\frac{3}{2}\right)+\frac{46}{3}\right)+\li3
   \left(-\frac{x^2}{3}
\brk
+\frac{23 x}{6}+\ly \left(-10 x^2-3
   x+\frac{15}{2}\right)+\lx \left(24 x^2-10
   x+5\right)+\frac{29}{6}\right)+\li2 \left(\left(-21 x^2+\frac{21
   x}{2}
\ibrk
-\frac{21}{4}\right) \lx^2+\left(\frac{x^2}{3}-\frac{23
   x}{6}+\ly \left(10 x^2+3 x-\frac{15}{2}\right)-\frac{29}{6}\right)
   \lx+\pi ^2 \left(-5
   x^2-\frac{x}{6}+\frac{25}{12}\right)\right)
\ebrk
+\left(\frac{43
   x^2}{18}+\frac{28 x}{9}-\frac{169}{18}\right) \z3+\ly
   \left(\frac{520 x^2}{9}-\frac{829 x}{18}+\pi ^2 \left(-\frac{50
   x^2}{9}+\frac{28 x}{3}-\frac{25}{3}\right)+\left(\frac{98 x^2}{3}
\ibrk
-\frac{107
   x}{3}+\frac{71}{6}\right) \z3+\frac{605}{36}\right)+\pi ^2
   \left(-\frac{3797 x^2}{216}+\frac{1585 x}{108}+\left(2 x^2-2 x+1\right)
   \log (2)-\frac{2353}{216}\right)
\ebrk
+\frac{41473}{1296},
\end{eqnarray}

\begin{eqnarray}
C &=&
\frac{1}{\epsilon^4}\left\{
\frac{x^2}{2}-\frac{x}{2}+\frac{1}{4}
\right\}
+
\frac{1}{\epsilon^3}\left\{
\lm\left[
x^2-x+\frac{1}{2}
\right]
+
\ls\left[
-x^2+x-\frac{1}{2}
\right]
+
\frac{5 x^2}{2}-\frac{5 x}{2}+\lx \left(-4 x^2+4 x
\ibrk
-2\right)+\ly
   \left(4 x^2-4 x+2\right)+1
\right\}
+
\frac{1}{\epsilon^2}\left\{
\lm \ls\left[
-2 x^2+2 x-1
\right]
+
\ls^2\left[
x^2-x+\frac{1}{2}
\right]
+
\lm\left[
3 x^2-3 x
\ibrk
+\lx \left(-4 x^2+4 x-2\right)+\ly \left(4 x^2-4
   x+2\right)+1
\right]
+
\ls\left[
-5 x^2+5 x+\ly \left(-8 x^2+8 x-4\right)+\lx \left(8 x^2
\iibrk
-8
   x+4\right)-2
\right]
+
\left(6 x^2-7 x+\frac{7}{2}\right) \lx^2+\left(-10 x^2+11 x+\ly
   \left(-12 x^2+12 x-6\right)-4\right) \lx+\frac{73 x^2}{8}
\brk
-\frac{73
   x}{8}+\pi ^2 \left(-\frac{13 x^2}{4}+\frac{13
   x}{4}-\frac{13}{8}\right)+\ly^2 \left(6 x^2-7
   x+\frac{7}{2}\right)+\ly \left(10 x^2-9 x+3\right)+\frac{53}{16}
\right\}
\ebrk
+
\frac{1}{\epsilon}\left\{
\lm^3\left[
-\frac{x^2}{3}+\frac{x}{3}-\frac{1}{6}
\right]
+
\lm \ls^2\left[
2 x^2-2 x+1
\right]
+
\ls^3\left[
-\frac{2 x^2}{3}+\frac{2 x}{3}-\frac{1}{3}
\right]
+
\lm^2\left[
-x^2+x
\ibrk
+\ly \left(-2 x^2+2 x-1\right)+\lx \left(2 x^2-2
   x+1\right)-\frac{1}{2}
\right]
+
\lm \ls\left[
-6 x^2+6 x+\ly \left(-8 x^2+8 x-4\right)
\ibrk
+\lx \left(8 x^2-8
   x+4\right)-2
\right]
+
\ls^2\left[
5 x^2-5 x+\lx \left(-8 x^2+8 x-4\right)+\ly \left(8 x^2-8
   x+4\right)+2
\right]
\brk
+
\lm\left[
\left(\frac{1}{2}-x\right) \lx^2+\left(3 x-2 x^2\right)
   \lx+\frac{37 x^2}{4}+\ly^2 \left(\frac{1}{2}-x\right)-\frac{37
   x}{4}+\pi ^2 \left(-\frac{7 x^2}{6}+\frac{7
   x}{6}-\frac{7}{12}\right)
\ibrk
+\ly \left(2 x^2-x-1\right)+\frac{25}{8}
\right]
+
\ls\left[
\left(-12 x^2+14 x-7\right) \lx^2+\left(20 x^2-22 x+\ly \left(24
   x^2-24 x+12\right)
\iibrk
+8\right) \lx-\frac{73 x^2}{4}+\frac{73
   x}{4}+\ly \left(-20 x^2+18 x-6\right)+\ly^2 \left(-12 x^2+14
   x-7\right)+\pi ^2 \left(\frac{13 x^2}{2}-\frac{13
   x}{2}
\iibrk
+\frac{13}{4}\right)-\frac{53}{8}
\right]
+
\left(\frac{2 x^2}{3}+3 x-\frac{3}{2}\right) \lx^3+\left(-6 x+\ly
   \left(-2 x^2-2 x+1\right)+\frac{3}{2}\right) \lx^2+\li2
   \left(-4 x^2
\ibrk
+2 x-1\right) \lx+\left(\left(3 x-\frac{3}{2}\right)
   \ly^2+3 \ly-24 x^2+\frac{55 x}{2}+\pi ^2 \left(\frac{46
   x^2}{3}-\frac{31 x}{3}+\frac{31}{6}\right)-10\right) \lx
\brk
-\frac{3
   \ly^2}{2}+\frac{429 x^2}{16}-\frac{429 x}{16}+\pi ^2 \left(-\frac{32
   x^2}{3}+\frac{17 x}{3}-\frac{29}{24}\right)+\ly^3 \left(-\frac{2
   x^2}{3}-\frac{5 x}{3}+\frac{5}{6}\right)+\s12 \left(4 x^2
\ibrk
-6
   x+3\right)+\li3 \left(4 x^2-2 x+1\right)+\ly \left(24
   x^2-\frac{41 x}{2}+\pi ^2 \left(-\frac{46 x^2}{3}+\frac{61
   x}{3}-\frac{61}{6}\right)+\frac{13}{2}\right)
\brk
+\left(-\frac{34
   x^2}{3}+\frac{28 x}{3}-\frac{14}{3}\right) \z3+\frac{283}{32}
\right\}
+
\lm^4\left[
\frac{x^2}{4}-\frac{x}{4}+\frac{1}{8}
\right]
+
\lm^3 \ls\left[
\frac{2 x^2}{3}-\frac{2 x}{3}+\frac{1}{3}
\right]
\ebrk
+
\lm \ls^3\left[
-\frac{4 x^2}{3}+\frac{4 x}{3}-\frac{2}{3}
\right]
+
\ls^4\left[
\frac{x^2}{3}-\frac{x}{3}+\frac{1}{6}
\right]
+
\lm^3\left[
\lx \left(-\frac{2 x^2}{3}+\frac{2 x}{3}-\frac{1}{3}\right)+\ly
   \left(\frac{2 x^2}{3}-\frac{2 x}{3}
\ibrk
+\frac{1}{3}\right)+\frac{1}{6}
\right]
+
\lm^2 \ls\left[
2 x^2-2 x+\lx \left(-4 x^2+4 x-2\right)+\ly \left(4 x^2-4
   x+2\right)+1
\right]
+
\lm \ls^2\left[
6 x^2
\brk
-6 x+\lx \left(-8 x^2+8 x-4\right)+\ly \left(8 x^2-8
   x+4\right)+2
\right]
+
\ls^3\left[
-\frac{10 x^2}{3}+\frac{10 x}{3}+\ly \left(-\frac{16 x^2}{3}+\frac{16
   x}{3}
\ibrk
-\frac{8}{3}\right)+\lx \left(\frac{16 x^2}{3}-\frac{16
   x}{3}+\frac{8}{3}\right)-\frac{4}{3}
\right]
+
\lm^2\left[
\left(\frac{x}{2}-\frac{1}{4}\right) \lx^2+\left(x^2-\frac{3
   x}{2}\right) \lx-\frac{5 x^2}{2}+\ly^2
   \left(\frac{x}{2}
\ibrk
-\frac{1}{4}\right)+\frac{5 x}{2}+\ly
   \left(-x^2+\frac{x}{2}+\frac{1}{2}\right)+\pi ^2 \left(\frac{2
   x^2}{3}-\frac{2 x}{3}+\frac{1}{3}\right)-\frac{3}{4}
\right]
+
\lm \ls\left[
(2 x-1) \lx^2+\left(4 x^2
\ibrk
-6 x\right) \lx-\frac{37 x^2}{2}+\frac{37
   x}{2}+\ly^2 (2 x-1)+\ly \left(-4 x^2+2 x+2\right)+\pi ^2
   \left(\frac{7 x^2}{3}-\frac{7 x}{3}+\frac{7}{6}\right)-\frac{25}{4}
\right]
\ebrk
+
\ls^2\left[
\left(12 x^2-14 x+7\right) \lx^2+\left(-20 x^2+22 x+\ly \left(-24
   x^2+24 x-12\right)-8\right) \lx+\frac{73 x^2}{4}-\frac{73 x}{4}
\brk
+\pi
   ^2 \left(-\frac{13 x^2}{2}+\frac{13 x}{2}-\frac{13}{4}\right)+\ly^2
   \left(12 x^2-14 x+7\right)+\ly \left(20 x^2-18
   x+6\right)+\frac{53}{8}
\right]
+
\lm\left[
\frac{2 x^2 \lx^3}{3}
\brk
+\left(-x+\ly \left(-2
   x^2+x-\frac{1}{2}\right)+\frac{1}{2}\right) \lx^2+\li2 \left(-4
   x^2+2 x-1\right) \lx+\left(-8 x^2+\frac{19 x}{2}+\pi ^2 \left(2
   x^2
\iibrk
-x+\frac{1}{2}\right)-4\right) \lx+\frac{189 x^2}{8}+\ly^2
   \left(\frac{1}{2}-x\right)-\frac{189 x}{8}+\pi ^2 \left(-\frac{5
   x^2}{3}+\frac{2 x}{3}+\frac{1}{4}\right)+\ly^3 \left(-\frac{2
   x^2}{3}+\frac{4 x}{3}
\ibrk
-\frac{2}{3}\right)+\s12 \left(4 x^2-6
   x+3\right)+\li3 \left(4 x^2-2 x+1\right)+\ly \left(8
   x^2-\frac{13 x}{2}+\pi ^2 \left(-2 x^2+3
   x
\iibrk
-\frac{3}{2}\right)+\frac{5}{2}\right)+\left(-\frac{44 x^2}{3}+\frac{38
   x}{3}-\frac{19}{3}\right) \z3+\frac{115}{16}
\right]
+
\ls\left[
\left(-\frac{4 x^2}{3}-6 x+3\right) \lx^3+\left(12 x+\ly \left(4
   x^2
\iibrk
+4 x-2\right)-3\right) \lx^2+\li2 \left(8 x^2-4 x+2\right)
   \lx+\left((3-6 x) \ly^2-6 \ly+48 x^2-55 x+\pi ^2
   \left(-\frac{92 x^2}{3}
\iibrk
+\frac{62 x}{3}-\frac{31}{3}\right)+20\right)
   \lx+3 \ly^2-\frac{429 x^2}{8}+\frac{429 x}{8}+\li3
   \left(-8 x^2+4 x-2\right)+\s12 \left(-8 x^2
\ibrk
+12 x-6\right)+\ly^3
   \left(\frac{4 x^2}{3}+\frac{10 x}{3}-\frac{5}{3}\right)+\pi ^2
   \left(\frac{64 x^2}{3}-\frac{34 x}{3}+\frac{29}{12}\right)+\ly
   \left(-48 x^2+41 x+\pi ^2 \left(\frac{92 x^2}{3}
\iibrk
-\frac{122
   x}{3}+\frac{61}{3}\right)-13\right)+\left(\frac{68 x^2}{3}-\frac{56
   x}{3}+\frac{28}{3}\right) \z3-\frac{283}{16}
\right]
+
\left(-3 x^2-\frac{x}{12}+\frac{1}{24}\right) \lx^4+\left(\frac{5
   x^2}{3}
\brk
+\frac{7 x}{3}+\ly \left(12 x^2-\frac{13
   x}{3}+\frac{13}{6}\right)-\frac{17}{12}\right) \lx^3+\left(\left(-7
   x^2+\frac{7 x}{2}-\frac{7}{4}\right) \ly^2+\left(-5
   x^2-\frac{1}{4}\right) \ly
\brk
-\frac{83 x}{4}+\pi ^2 \left(-\frac{58
   x^2}{3}+\frac{34
   x}{3}-\frac{17}{3}\right)+\frac{3}{1-x}+\frac{35}{8}\right)
   \lx^2+\left(\left(2 x^2-4 x+2\right)
   \ly^3+\left(\frac{3}{4}-\frac{3 x}{2}\right) \ly^2
\brk
+\left(\pi ^2
   \left(\frac{116 x^2}{3}-\frac{89 x}{3}+\frac{89}{6}\right)-6\right)
   \ly-48 x^2+\frac{125 x}{2}+\pi ^2 \left(25 x^2-13
   x+\frac{7}{6}\right)+\left(\frac{128 x^2}{3}
\ibrk
-\frac{164
   x}{3}+\frac{82}{3}\right) \z3-\frac{97}{4}\right)
   \lx+\frac{2479 x^2}{32}-\frac{2479 x}{32}+\s13 \left(-36 x^2+62
   x-31\right)+\ly^4 \left(-3 x^2
\brk
+\frac{71
   x}{12}-\frac{71}{24}\right)+\ly^3 \left(-\frac{5 x^2}{3}+\frac{2
   x}{3}+\frac{11}{12}\right)+\pi ^4 \left(-\frac{829 x^2}{720}-\frac{191
   x}{720}+\frac{191}{1440}\right)+\s22 \left(8 x^2
\brk
-56
   x+28\right)+\li4 \left(12 x^2-14 x+7\right)+\ly^2
   \left(\frac{37 x}{4}+\pi ^2 \left(-\frac{20 x^2}{3}+\frac{29
   x}{3}-\frac{29}{6}\right)-\frac{37}{8}+\frac{3}{x}\right)
\ebrk
+\s12
   \left(10 x^2-11 x+\lx \left(-16 x^2+56 x-28\right)+\ly \left(8
   x^2-8 x+4\right)\right)+\li3 \left(10 x^2+3 x
\brk
+\lx \left(-28
   x^2+18 x-9\right)+\ly \left(16 x^2-20 x+10\right)-1\right)+\li2
   \left(\left(22 x^2-11 x+\frac{11}{2}\right) \lx^2
\brk
+\left(-10 x^2-3
   x+\ly \left(-16 x^2+20 x-10\right)+1\right) \lx+\pi ^2
   \left(\frac{76 x^2}{3}-\frac{52
   x}{3}+\frac{26}{3}\right)\right)+\left(-\frac{265 x^2}{6}
\brk
+\frac{259
   x}{6}-\frac{293}{12}\right) \z3+\ly \left(48 x^2-\frac{67
   x}{2}+\pi ^2 \left(-25 x^2+30 x-\frac{61}{6}\right)+\left(-\frac{128
   x^2}{3}+\frac{140 x}{3}
\ibrk
-\frac{70}{3}\right)
   \z3+\frac{39}{4}\right)+\pi ^2 \left(-\frac{1631 x^2}{48}+\frac{285
   x}{16}+\left(-2 x^2+2 x-1\right) \log
   (2)-\frac{247}{96}\right)+\frac{1621}{64},
\end{eqnarray}

\begin{eqnarray}
D_l &=&
\frac{1}{\epsilon^3}\left\{
-\frac{x^2}{2}+\frac{x}{2}-\frac{1}{4}
\right\}
+
\frac{1}{\epsilon^2}\left\{
\lm\left[
-\frac{x^2}{3}+\frac{x}{3}-\frac{1}{6}
\right]
+
\ls\left[
-\frac{x^2}{3}+\frac{x}{3}-\frac{1}{6}
\right]
+
\frac{5 x^2}{9}-\frac{5 x}{9}+\lx \left(\frac{2 x^2}{3}
\ibrk
-\frac{2
   x}{3}+\frac{1}{3}\right)+\frac{19}{36}
\right\}
+
\frac{1}{\epsilon}\left\{
\lm \ls\left[
-\frac{2 x^2}{3}+\frac{2 x}{3}-\frac{1}{3}
\right]
+
\ls^2\left[
x^2-x+\frac{1}{2}
\right]
+
\lm\left[
\frac{5 x^2}{3}-\frac{5 x}{3}+1
\right]
\brk
+
\ls\left[
-\frac{40 x^2}{9}+\frac{40 x}{9}+\lx \left(\frac{4 x^2}{3}-\frac{4
   x}{3}+\frac{2}{3}\right)-\frac{37}{18}
\right]
+
\frac{649 x^2}{108}-\frac{649 x}{108}+\lx \left(-\frac{10
   x^2}{3}+\frac{10 x}{3}-2\right)
\brk
+\pi ^2 \left(-\frac{11 x^2}{36}+\frac{11
   x}{36}-\frac{11}{72}\right)+\frac{589}{216}
\right\}
+
\lm^3\left[
\frac{x^2}{9}-\frac{x}{9}+\frac{1}{18}
\right]
+
\lm^2 \ls\left[
\frac{2 x^2}{3}-\frac{2 x}{3}+\frac{1}{3}
\right]
\ebrk
+
\lm \ls^2\left[
\frac{4 x^2}{3}-\frac{4 x}{3}+\frac{2}{3}
\right]
+
\ls^3\left[
-\frac{8 x^2}{9}+\frac{8 x}{9}-\frac{4}{9}
\right]
+
\lm^2\left[
-\frac{23 x^2}{18}+\frac{23 x}{18}-\frac{23}{36}
\right]
+
\lm \ls\left[
-4 x^2
\brk
+4 x-\frac{5}{3}
\right]
+
\ls^2\left[
\frac{11 x^2}{9}-\frac{11 x}{9}+\lx \left(-\frac{8 x^2}{3}+\frac{8
   x}{3}-\frac{4}{3}\right)+\frac{1}{9}
\right]
+
\lm\left[
\frac{73 x^2}{18}-\frac{73 x}{18}+\pi ^2
   \left(-\frac{x^2}{6}
\ibrk
+\frac{x}{6}-\frac{1}{12}\right)+\frac{43}{36}
\right]
+
\ls\left[
\left(\frac{2 x}{3}-\frac{1}{3}\right) \lx^2+\left(\frac{20
   x^2}{3}-\frac{22 x}{3}+\frac{10}{3}\right) \lx+\frac{91
   x^2}{18}-\frac{91 x}{18}+\pi ^2 \left(\frac{17 x^2}{18}
\ibrk
-\frac{17
   x}{18}+\frac{17}{36}\right)+\frac{19}{4}
\right]
+
\frac{2 x^2 \lx^3}{9}+\left(-\frac{13 x}{9}+\ly \left(-\frac{2
   x^2}{3}+\frac{2 x}{3}-\frac{1}{3}\right)+\frac{5}{9}\right)
   \lx^2+\li2 \left(-\frac{4 x^2}{3}
\brk
+\frac{4
   x}{3}-\frac{2}{3}\right) \lx+\left(-\frac{56 x^2}{9}+8 x+\pi ^2
   \left(\frac{25 x^2}{9}-\frac{17
   x}{9}+\frac{17}{18}\right)-\frac{29}{9}\right) \lx-\frac{2639
   x^2}{216}+\frac{2639 x}{216}
\ebrk
+\pi ^2 \left(\frac{2 x^2}{27}-\frac{14
   x}{27}+\frac{137}{216}\right)+\li3 \left(\frac{4 x^2}{3}-\frac{4
   x}{3}+\frac{2}{3}\right)+\left(-\frac{73 x^2}{9}+\frac{73
   x}{9}-\frac{73}{18}\right) \z3-\frac{3937}{432}, \ebrk
\end{eqnarray}

\begin{eqnarray}
D_h &=&
\frac{1}{\epsilon^2}\left\{
\lm\left[
-\frac{2 x^2}{3}+\frac{2 x}{3}-\frac{1}{3}
\right]
+
\ls\left[
-\frac{4 x^2}{3}+\frac{4 x}{3}-\frac{2}{3}
\right]
+
\frac{10 x^2}{9}-\frac{10 x}{9}+\frac{5}{9}
\right\}
+
\frac{1}{\epsilon}\left\{
\lm^2\left[
-\frac{x^2}{3}+\frac{x}{3}
\ibrk
-\frac{1}{6}
\right]
+
\ls^2\left[
2 x^2-2 x+1
\right]
+
\lm\left[
-\frac{5 x^2}{9}+\frac{5 x}{9}+\lx \left(\frac{4 x^2}{3}-\frac{4
   x}{3}+\frac{2}{3}\right)+\frac{1}{18}
\right]
+
\ls\left[
-\frac{50 x^2}{9}
\ibrk
+\frac{50 x}{9}+\lx \left(\frac{8 x^2}{3}-\frac{8
   x}{3}+\frac{4}{3}\right)-\frac{19}{9}
\right]
+
\frac{131 x^2}{27}-\frac{131 x}{27}+\lx \left(-\frac{20 x^2}{9}+\frac{20
   x}{9}-\frac{10}{9}\right)
\brk
+\pi ^2
   \left(-\frac{x^2}{3}+\frac{x}{3}-\frac{1}{6}\right)+\frac{101}{54}
\right\}
+
\lm^3\left[
\frac{7 x^2}{9}-\frac{7 x}{9}+\frac{7}{18}
\right]
+
\lm^2 \ls\left[
\frac{4 x^2}{3}-\frac{4 x}{3}+\frac{2}{3}
\right]
+
\lm \ls^2\left[
\frac{2 x^2}{3}
\brk
-\frac{2 x}{3}+\frac{1}{3}
\right]
+
\ls^3\left[
-\frac{14 x^2}{9}+\frac{14 x}{9}-\frac{7}{9}
\right]
+
\lm^2\left[
\frac{37 x^2}{18}-\frac{37 x}{18}+\lx \left(-\frac{4 x^2}{3}+\frac{4
   x}{3}-\frac{2}{3}\right)+\frac{43}{36}
\right]
\ebrk
+
\lm \ls\left[
\frac{4 x^2}{9}-\frac{4 x}{9}+\lx \left(-\frac{8 x^2}{3}+\frac{8
   x}{3}-\frac{4}{3}\right)+\frac{2}{9}
\right]
+
\ls^2\left[
\frac{7 x^2}{3}-\frac{7 x}{3}+\lx \left(-4 x^2+4 x-2\right)+\frac{1}{6}
\right]
\ebrk
+
\lm\left[
\frac{76 x^2}{9}-\frac{76 x}{9}+\lx \left(-\frac{20 x^2}{9}+\frac{20
   x}{9}-\frac{16}{9}\right)+\frac{9}{2}
\right]
+
\ls\left[
\left(\frac{2 x}{3}-\frac{1}{3}\right) \lx^2+\left(\frac{40
   x^2}{9}-\frac{46 x}{9}
\ibrk
+\frac{14}{9}\right) \lx+\frac{199
   x^2}{27}-\frac{199 x}{27}+\pi ^2
   \left(x^2-x+\frac{1}{2}\right)+\frac{349}{54}
\right]
+
\frac{2 x^2 \lx^3}{9}+\left(-\frac{13 x}{9}+\ly \left(-\frac{2
   x^2}{3}+\frac{2 x}{3}
\ibrk
-\frac{1}{3}\right)+\frac{5}{9}\right)
   \lx^2+\li2 \left(-\frac{4 x^2}{3}+\frac{4
   x}{3}-\frac{2}{3}\right) \lx+\left(-\frac{224 x^2}{27}+\frac{272
   x}{27}+\pi ^2 \left(\frac{8 x^2}{3}-\frac{16
   x}{9}+\frac{8}{9}\right)
\brk
-\frac{130}{27}\right) \lx+\frac{113
   x^2}{54}-\frac{113 x}{54}+\pi ^2 \left(\frac{17 x^2}{54}-\frac{41
   x}{54}+\frac{83}{108}\right)+\li3 \left(\frac{4 x^2}{3}-\frac{4
   x}{3}+\frac{2}{3}\right)
\ebrk
+\left(-\frac{76 x^2}{9}+\frac{76
   x}{9}-\frac{38}{9}\right) \z3-\frac{149}{108},
\end{eqnarray}

\begin{eqnarray}
E_l &=&
\frac{1}{\epsilon^3}\left\{
\frac{x^2}{2}-\frac{x}{2}+\frac{1}{4}
\right\}
+
\frac{1}{\epsilon^2}\left\{
\lm\left[
\frac{x^2}{3}-\frac{x}{3}+\frac{1}{6}
\right]
+
\ls\left[
\frac{x^2}{3}-\frac{x}{3}+\frac{1}{6}
\right]
-\frac{5 x^2}{9}+\frac{5 x}{9}+\lx \left(-\frac{4 x^2}{3}+\frac{4
   x}{3}
\ibrk
-\frac{2}{3}\right)+\ly \left(\frac{4 x^2}{3}-\frac{4
   x}{3}+\frac{2}{3}\right)-\frac{19}{36}
\right\}
+
\frac{1}{\epsilon}\left\{
\lm \ls\left[
\frac{2 x^2}{3}-\frac{2 x}{3}+\frac{1}{3}
\right]
+
\ls^2\left[
-x^2+x-\frac{1}{2}
\right]
\brk
+
\lm\left[
-\frac{5 x^2}{3}+\frac{5 x}{3}-1
\right]
+
\ls\left[
\frac{40 x^2}{9}-\frac{40 x}{9}+\lx \left(-\frac{8 x^2}{3}+\frac{8
   x}{3}-\frac{4}{3}\right)+\ly \left(\frac{8 x^2}{3}-\frac{8
   x}{3}+\frac{4}{3}\right)
\ibrk
+\frac{37}{18}
\right]
-\frac{649 x^2}{108}+\frac{649 x}{108}+\ly \left(-\frac{20
   x^2}{3}+\frac{20 x}{3}-4\right)+\pi ^2 \left(\frac{59 x^2}{36}-\frac{59
   x}{36}+\frac{59}{72}\right)+\lx \left(\frac{20 x^2}{3}
\ibrk
-\frac{20
   x}{3}+4\right)-\frac{589}{216}
\right\}
+
\lm^3\left[
-\frac{x^2}{9}+\frac{x}{9}-\frac{1}{18}
\right]
+
\lm^2 \ls\left[
-\frac{2 x^2}{3}+\frac{2 x}{3}-\frac{1}{3}
\right]
+
\lm \ls^2\left[
-\frac{4 x^2}{3}+\frac{4 x}{3}
\brk
-\frac{2}{3}
\right]
+
\ls^3\left[
\frac{8 x^2}{9}-\frac{8 x}{9}+\frac{4}{9}
\right]
+
\lm^2\left[
\frac{23 x^2}{18}-\frac{23 x}{18}+\frac{23}{36}
\right]
+
\lm \ls\left[
4 x^2-4 x+\frac{5}{3}
\right]
+
\ls^2\left[
-\frac{55 x^2}{9}
\brk
+\frac{55 x}{9}+\ly \left(-\frac{16 x^2}{3}+\frac{16
   x}{3}-\frac{8}{3}\right)+\lx \left(\frac{16 x^2}{3}-\frac{16
   x}{3}+\frac{8}{3}\right)-\frac{23}{9}
\right]
+
\lm\left[
-\frac{73 x^2}{18}+\frac{73 x}{18}
\brk
+\pi ^2
   \left(\frac{x^2}{6}-\frac{x}{6}+\frac{1}{12}\right)-\frac{43}{36}
\right]
+
\ls\left[
\left(\frac{2}{3}-\frac{4 x}{3}\right) \lx^2+\left(-\frac{40
   x^2}{3}+\frac{44 x}{3}-\frac{20}{3}\right) \lx+\frac{1027
   x^2}{54}
\brk
+\ly^2 \left(\frac{2}{3}-\frac{4 x}{3}\right)-\frac{1027
   x}{54}+\pi ^2 \left(-\frac{89 x^2}{18}+\frac{89
   x}{18}-\frac{89}{36}\right)+\ly \left(\frac{40 x^2}{3}-12
   x+\frac{16}{3}\right)+\frac{787}{108}
\right]
\ebrk
-\frac{4}{9} x^2 \lx^3+\left(\frac{26 x}{9}+\ly \left(\frac{4
   x^2}{3}-\frac{4 x}{3}+\frac{2}{3}\right)-\frac{10}{9}\right)
   \lx^2+\li2 \left(\frac{8 x^2}{3}-\frac{8
   x}{3}+\frac{4}{3}\right) \lx+\left(\frac{112 x^2}{9}
\brk
-16 x+\pi ^2
   \left(-\frac{50 x^2}{9}+\frac{34
   x}{9}-\frac{17}{9}\right)+\frac{58}{9}\right) \lx-\frac{13963
   x^2}{648}+\frac{13963 x}{648}+\ly^2 \left(\frac{26
   x}{9}-\frac{16}{9}\right)
\ebrk
+\li3 \left(-\frac{8 x^2}{3}+\frac{8
   x}{3}-\frac{4}{3}\right)+\s12 \left(-\frac{8 x^2}{3}+\frac{8
   x}{3}-\frac{4}{3}\right)+\ly^3 \left(\frac{4 x^2}{9}-\frac{8
   x}{9}+\frac{4}{9}\right)
\ebrk
+\pi ^2 \left(\frac{250 x^2}{27}-\frac{202
   x}{27}+\frac{631}{216}\right)+\ly \left(-\frac{112 x^2}{9}+\frac{80
   x}{9}+\pi ^2 \left(\frac{50 x^2}{9}-\frac{22
   x}{3}+\frac{11}{3}\right)-\frac{26}{9}\right)
\ebrk
+\left(\frac{13
   x^2}{9}-\frac{13 x}{9}+\frac{13}{18}\right) \z3-\frac{10069}{1296},
\end{eqnarray}

\begin{eqnarray}
E_h &=&
\frac{1}{\epsilon^2}\left\{
\lm\left[
\frac{2 x^2}{3}-\frac{2 x}{3}+\frac{1}{3}
\right]
+
\ls\left[
\frac{4 x^2}{3}-\frac{4 x}{3}+\frac{2}{3}
\right]
+
-\frac{10 x^2}{9}+\frac{10 x}{9}-\frac{5}{9}
\right\}
+
\frac{1}{\epsilon}\left\{
\lm^2\left[
\frac{x^2}{3}-\frac{x}{3}+\frac{1}{6}
\right]
\brk
+
\ls^2\left[
-2 x^2+2 x-1
\right]
+
\lm\left[
\frac{5 x^2}{9}-\frac{5 x}{9}+\lx \left(-\frac{8 x^2}{3}+\frac{8
   x}{3}-\frac{4}{3}\right)+\ly \left(\frac{8 x^2}{3}-\frac{8
   x}{3}+\frac{4}{3}\right)-\frac{1}{18}
\right]
\brk
+
\ls\left[
\frac{50 x^2}{9}-\frac{50 x}{9}+\lx \left(-\frac{16 x^2}{3}+\frac{16
   x}{3}-\frac{8}{3}\right)+\ly \left(\frac{16 x^2}{3}-\frac{16
   x}{3}+\frac{8}{3}\right)+\frac{19}{9}
\right]
-\frac{131 x^2}{27}
\brk
+\frac{131 x}{27}+\ly \left(-\frac{40
   x^2}{9}+\frac{40 x}{9}-\frac{20}{9}\right)+\pi ^2 \left(\frac{5
   x^2}{3}-\frac{5 x}{3}+\frac{5}{6}\right)+\lx \left(\frac{40
   x^2}{9}-\frac{40 x}{9}+\frac{20}{9}\right)
\brk
-\frac{101}{54}
\right\}
+
\lm^3\left[
-\frac{7 x^2}{9}+\frac{7 x}{9}-\frac{7}{18}
\right]
+
\lm^2 \ls\left[
-\frac{4 x^2}{3}+\frac{4 x}{3}-\frac{2}{3}
\right]
+
\lm \ls^2\left[
-\frac{2 x^2}{3}+\frac{2 x}{3}-\frac{1}{3}
\right]
\ebrk
+
\ls^3\left[
\frac{14 x^2}{9}-\frac{14 x}{9}+\frac{7}{9}
\right]
+
\lm^2\left[
-\frac{37 x^2}{18}+\frac{37 x}{18}+\ly \left(-\frac{8 x^2}{3}+\frac{8
   x}{3}-\frac{4}{3}\right)+\lx \left(\frac{8 x^2}{3}-\frac{8
   x}{3}+\frac{4}{3}\right)
\brk
-\frac{43}{36}
\right]
+
\lm \ls\left[
-\frac{4 x^2}{9}+\frac{4 x}{9}+\ly \left(-\frac{16 x^2}{3}+\frac{16
   x}{3}-\frac{8}{3}\right)+\lx \left(\frac{16 x^2}{3}-\frac{16
   x}{3}+\frac{8}{3}\right)-\frac{2}{9}
\right]
\ebrk
+
\ls^2\left[
-\frac{65 x^2}{9}+\frac{65 x}{9}+\ly \left(-8 x^2+8 x-4\right)+\lx
   \left(8 x^2-8 x+4\right)-\frac{47}{18}
\right]
+
\lm\left[
-\frac{76 x^2}{9}+\frac{76 x}{9}
\brk
+\ly \left(-\frac{40 x^2}{9}+\frac{40
   x}{9}-\frac{32}{9}\right)+\lx \left(\frac{40 x^2}{9}-\frac{40
   x}{9}+\frac{32}{9}\right)-\frac{9}{2}
\right]
+
\ls\left[
\left(\frac{2}{3}-\frac{4 x}{3}\right) \lx^2+\left(-\frac{80
   x^2}{9}
\ibrk
+\frac{92 x}{9}-\frac{28}{9}\right) \lx+\frac{451
   x^2}{27}+\ly^2 \left(\frac{2}{3}-\frac{4 x}{3}\right)-\frac{451
   x}{27}+\pi ^2 \left(-5 x^2+5 x-\frac{5}{2}\right)+\ly \left(\frac{80
   x^2}{9}
\ibrk
-\frac{68 x}{9}+\frac{16}{9}\right)+\frac{301}{54}
\right]
+
-\frac{4}{9} x^2 \lx^3+\left(\frac{26 x}{9}+\ly \left(\frac{4
   x^2}{3}-\frac{4 x}{3}+\frac{2}{3}\right)-\frac{10}{9}\right)
   \lx^2+\li2 \left(\frac{8 x^2}{3}
\brk
-\frac{8
   x}{3}+\frac{4}{3}\right) \lx+\left(\frac{448 x^2}{27}-\frac{544
   x}{27}+\pi ^2 \left(-\frac{16 x^2}{3}+\frac{32
   x}{9}-\frac{16}{9}\right)+\frac{260}{27}\right) \lx-\frac{5809
   x^2}{162}
\ebrk
+\frac{5809 x}{162}+\ly^2 \left(\frac{26
   x}{9}-\frac{16}{9}\right)+\li3 \left(-\frac{8 x^2}{3}+\frac{8
   x}{3}-\frac{4}{3}\right)+\s12 \left(-\frac{8 x^2}{3}+\frac{8
   x}{3}-\frac{4}{3}\right)
\ebrk
+\ly^3 \left(\frac{4 x^2}{9}-\frac{8
   x}{9}+\frac{4}{9}\right)+\pi ^2 \left(\frac{487 x^2}{54}-\frac{391
   x}{54}+\frac{301}{108}\right)+\ly \left(-\frac{448 x^2}{27}+\frac{352
   x}{27}+\pi ^2 \left(\frac{16 x^2}{3}
\ibrk
-\frac{64
   x}{9}+\frac{32}{9}\right)-\frac{164}{27}\right)+\left(\frac{16
   x^2}{9}-\frac{16 x}{9}+\frac{8}{9}\right) \z3-\frac{5023}{324},
\end{eqnarray}

\begin{eqnarray}
F &=&
\ls^2\left[
\frac{4 x^2}{9}-\frac{4 x}{9}+\frac{2}{9}
\right]
+
\ls\left[
-\frac{40 x^2}{27}+\frac{40 x}{27}-\frac{20}{27}
\right]
+
\frac{100 x^2}{81}-\frac{100 x}{81}+\pi ^2 \left(-\frac{4 x^2}{9}+\frac{4
   x}{9}-\frac{2}{9}\right)+\frac{50}{81}. \ebrk
\end{eqnarray}
%%STOP
%
}}
Notice that our results are expressed not only in terms of the classic
polylogarithms up to weight four, but also in terms of Nielsen polylogarithms
\begin{equation}
\mathrm{S}_{n,p}(x) = \frac{(-1)^{n+p-1}}{(n-1)!p!} \int_0^1 dy
\frac{\log^{n-1}(y)\log^p(1-x y)}{y}.
\end{equation}
%

%
% -----------------------------------------------------------------------------
%
\subsection*{Conclusions}

In this letter, we have presented the two-loop virtual QCD corrections to the
production of heavy-quarks in the light quark annihilation channel in the
ultra-relativistic limit. Our results form a crucial part of the NNLO predictions for
heavy-quark production in hadron-hadron collisions. However, depending on the
mass of the heavy quark (bottom or top) and the kinematics of the process
under consideration power corrections in the heavy-quark mass may have to be 
considered as well.
Our results have been derived by combining two completely different methods which have substantial overlap. 
This provides direct and highly non-trivial checks of the QCD
factorization approach of Ref.~\cite{Mitov:2006xs}, 
of the direct calculation of massive Feynman diagrams and, last but not least,
also on the available masseless results of Ref.~\cite{Anastasiou:2000kg}.

In order to yield physical cross sections, our result for $\langle {\cal M}^{(0)} | {\cal M}^{(2)} \rangle $
still has to be combined with the tree-level $2 \to 4$, the one-loop $2 \to 3$
as well as the square of the one-loop $2 \to 2$ processes.
While some of the matrix elements (including the full mass dependence)
can be easily generated, others became available in the literature only rather recently,
see e.g. Refs.\cite{Korner:2005rg,Dittmaier:2007wz}.
The combination of all these contributions enables the analytic cancellation of the remaining
infrared divergences as well as the isolation of the initial state singularities
which need to be absorbed into parton distribution functions.
This is a necessary prerequisite e.g. to the construction of numerical programs
which provide NNLO QCD estimates of observable scattering cross sections.

Finally, while we have chosen to work in this letter with squared matrix elements,
it should, of course, be clear that with the help of Refs.~\cite{DeFreitas:2004tk,Glover:2004si}
analogous results can be derived for the massive two-loop amplitude
$|{\cal M}^{(2)} \rangle$ itself.

A {\sc Mathematica} file with our results can be obtained by downloading the
source from the preprint server {\tt http://arXiv.org}. 
The results are also available from the authors upon request.

%
% ---------------------------------------------------------------------
%
{\bf{Acknowledgments:}}
We acknowledge useful discussions with
N.~Glover, J.~Gluza, Z.~Merebashvili and T.~Riemann.
The work of M.C. was supported by the Sofja Kovalevskaja Award of the 
Alexander von Humboldt Foundation 
and by the ToK Program ``ALGOTOOLS'' (MTKD-CD-2004-014319).
A.M. thanks the Alexander von Humboldt Foundation for support and
S.M. acknowledges contract VH-NG-105 of the Helmholtz Gemeinschaft.
This work was supported in part by the Deutsche Forschungsgemeinschaft in
Sonderforschungs\-be\-reich/Transregio~9.

%
% -----------------------------------------------------------------------------
%

\end{document}